# Geometrized Symbols and the Related Codes

Shahid Nawaz

*Department of Physics, University at Albany, State University of New York, NY 12222*

Email: sn439165@albany.edu

## Abstract

In this paper geometry is studied with a novel approach. Every geometrical object is defined as a symbol which satisfies some properties. These symbols are then coded into a class of numbers which are named here as '*many dots numbers* (MDN)'. The algebraic structure of MDN is established. Assuming the universe as a symbol, the existence of dark matter is explained qualitatively. Moreover, it is shown that dark matter increases as the universe expands.

## 1. Introduction

Real numbers play a key role in general relativity (GR) while complex numbers are central in quantum mechanics (QM). One may ask if there exists $i = \sqrt{-1}$ in QM then why it does not appear in GR [1]. This mystery in QM has been addressed by A. Caticha [2, 3] and P. Goyal, K. Knuth and J. Skilling (GKS) [4].

General relativity is in fact geometrodynamics. It is expected that geometry may play a central role in the unified theory. Therefore based on geometry, this paper proposes another class of numbers which may provide a unified description of QM and GR.

This paper is organized as follow. The model is presented in section-2. The results are qualitatively discussed in section-3. The paper is concluded in section-4.

## 2. Model

A symbol is defined as;

**Definition 2.1:** *A symbol is a geometrical object which has the following properties*:

*R1:     A symbol is a collection of geodesic segments of the hypersurface on which it is drawn.*

*R2:     Every symbol is connected.*

*R3:     Every symbol tells us how the previous and the following symbols look like.*

*R4:     To draw a symbol on d-dimensional hypersurface one needs to have 2d-arms.*

*R5:     The first symbol is a dot. A dot is a unit pixel.*

*R6:     The next symbol is drawn by attaching dots symmetrically in all degree of freedom keeping the first symbol in the center.*

In general one can draw any symbol which does not necessarily posses the rules required by definition-2.1. Such symbols will be called broken symbols. Broken symbol is defined below.

**Definition 2.2:** *A symbol is said to be broken if it does not possess some properties required by definition-2.1. A symbol is said to be maximally broken if it possesses no property required by definition-2.1.*

## 2.1. Symbols of flat space

We construct symbols which satisfy definition-2.1. Broken symbols will be discussed in section-2.3. Consider one dimensional case. We start with simple symbols which qualify the criteria of by definion-2.1. Draw a square box and color it dark to get the first symbol. In this case a squire box is considered to be a unit pixel, as shown in figure (2.1).

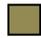

Figure 2.1: First symbol of one-dimension.

To draw the next symbol, concatenate the same size white square boxes on both sides of the box in figure (2.1). We want the center with a different color which is helpful to recognize the first symbol. The second symbol is shown in figure (2.2). Similarly the third symbol can be drawn by extending the second symbol by concatenating two white color boxes on both side of the second symbol and so on.

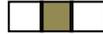

Figure (2.2): Second symbol in one dimension.

The symbols shown in figure (2.1) and figure (2.2) are coded, as shown in figure (2.3). The first symbol is coded into •. The first symbol remains in the center of every other symbol. In all the following symbols, the horizontal white boxes (one-dimensional case) are represented by **1**. The reason is that each point of, say *x*-axis, remains a point of the same axis. The *x*-axis is represented by **1**-axis**.** The origin of **1**-axis is represented by •. The strings of **1**'s to the left and right of the dot represent the negative and positive *x*-axes respectively. If the string is finite then it represents a finite segment of *x*-axis. The codeword **1•1** means that • has one space point to the left and one to the right. The codeword **11•11** can be similarly understood. The codeword $(\bar{\mathbf{1}} \cdot \bar{\mathbf{1}})_l$ means two strings of **1**'s each of length *l* is concatenated to the left and right of •. It represents a finite **1**-axis. An infinite **1**-axis is represented by $(\bar{\mathbf{1}} \cdot \bar{\mathbf{1}})_\infty \stackrel{\text{def}}{=} \bar{\mathbf{1}} \cdot \bar{\mathbf{1}}$. Said differently, the strings of **1**'s to the left and right of the dot represent the left and right arms respectively and the dot in the middle of the strings represents shaking of the two hands. In similar fashion **2**-, **3**-,…, and *d*-axis can be constructed. For instant, *d*-axis will be represented by $\bar{\mathbf{d}} \cdot \bar{\mathbf{d}}$. .

    ■ = •

    □■□ = **1•1**

    □□■□□ = **11•11**

Figure (2.3): The first, second and third symbols are coded.

The length of $(\bar{1} \bullet \bar{1})_l$ is denoted by $\left|(\bar{1} \bullet \bar{1})_l\right|_s$. Since the number of boxes are odd irrespective of their colors. Therefore

$$\left|(\bar{1} \bullet \bar{1})_l\right|_s = (2l+1)_s, \qquad l = 0, 1, 2, \ldots$$

The subscript $s$ is the scaling index. $(2l+1)_s$ is the odd number analog of real numbers.

$$\left\{\left|(\bar{1} \bullet \bar{1})_l\right|_s\right\} = \{1_s, 3_s, 5_s, \ldots, (2l+1)_s\}$$

For $l = 0$, $|\bullet|_s = 1_s$, also $1_{s_1} \neq 1_{s_2}$, for $s_1 \neq s_1$. For example, 1 m $\neq$ 1 cm. In our notations 1 m is written as $1_m$. The scaling index will be calculated in section-2.2.

In the above symbols we have chosen a dark color for the central box and white color for the boxes concatenated to it. We introduce color operator $\hat{C}$ which takes • to **1** and vice versa;

$$\hat{C}(\bar{1} \bullet \bar{1})_l = (\bar{\bullet} \mathbf{1} \bar{\bullet})_l$$

It is easy to check

$$\hat{C}^2(\bar{1} \bullet \bar{1})_l = (\bar{1} \bullet \bar{1})_l$$

and

$$\left|\hat{C}(\bar{1} \bullet \bar{1})_l\right|_s = \left|(\bar{1} \bullet \bar{1})_l\right|_s$$

It is important to give these numbers a name. The numbers described above may have more than one dots (see for instant, $(\bar{\bullet} \mathbf{1} \bar{\bullet})_l$ above). We shall call it '*many dots numbers* (MDN)' which means that unlike real numbers MDN may have more than one dots. The set of MDN is denoted by $\mathcal{M}$ and the set of corresponding symbols is denoted by $\mathfrak{H}$.

$$\mathcal{M} = \{\bullet, \mathbf{1} \bullet \mathbf{1}, \mathbf{11} \bullet \mathbf{11}, \ldots\}$$

The map $f: \mathfrak{H} \to \mathcal{M}$ takes a symbol into its code. Further the map $|\ldots|_s: \mathcal{M} \to O_s$ gives length of an element of $\mathcal{M}$.

$$O_s = \{1_s, 3_s, 5_s, \ldots\}$$

We now define addition, subtraction, multiplication and quotient rules of MDN.

**Addition:** Define an operation $\oplus$ such that

$$\mathbf{1} \bullet \mathbf{1} = \mathbf{1} \oplus \bullet \oplus \mathbf{1}$$

The left and right $\oplus$ mean to concatenate the left and right numbers (in this case **1**) to •. It should be noted that $\mathbf{1} \oplus \mathbf{1} \oplus \bullet$ is undefined. If it exists then that would correspond $\mathbf{11}\bullet$ which is not possible by definition-2.1. The later number corresponds to a broken symbol.

**Subtraction:** Define an operation $\ominus$ such that

$$\bullet = \mathbf{1} \ominus \mathbf{1} \bullet \mathbf{1} \ominus \mathbf{1}$$

Here the left and right $\ominus$ mean to remove **1**'s from the number in the middle. By a similar argument as in the addition rules, $\mathbf{1} \ominus \mathbf{1}$ is undefined.

**Multiplication:** To define multiplication, first we need to draw a two dimensional symbol.

$$\boxed{\phantom{x}} = (\mathbf{1} \otimes \mathbf{2})_1 = \mathbf{21} \bullet \mathbf{12}$$

Figure (2.4): The second symbol of two-dimensional space.

In figure (2.4), the second symbol of a two dimensional Euclidean space is drawn. It should be noted that the first symbol is always a dot by definition-2.1. The subscript 1 means that one white box is concatenated to the dark box along all possible directions. If $l$ boxes are concatenated to the dark box along all possible directions, then $(\mathbf{1} \otimes \mathbf{2})_l = (\overline{\mathbf{21}} \bullet \overline{\mathbf{12}})_l$. Here the hands of left, right, top and bottom arms are shacked at the central dark box. The infinite two dimensional axes are represented by $(\mathbf{1} \otimes \mathbf{2})_\infty = \overline{\mathbf{21}} \bullet \overline{\mathbf{12}}$.

One can count the total number of boxes in two-dimensional number;

$$|(\overline{\mathbf{21}} \bullet \overline{\mathbf{12}})_l|_{s_2} = (2^2 l + 1)_{s_2}$$

In $d$-dimensions

$$(\mathbf{1} \otimes \mathbf{2} \otimes \ldots \otimes \mathbf{d})_l = (\overline{\mathbf{d} \ldots \mathbf{21}} \bullet \overline{\mathbf{12} \ldots \mathbf{d}})_l$$

$$|(\mathbf{1} \otimes \mathbf{2} \otimes \ldots \otimes \mathbf{d})_l|_{s_d} = (2^d l + 1)_{s_d}, \quad l = 0,1,2,\ldots \text{ and } d = 1,2,3,\ldots$$

where $s_d$ is the scaling index.

**Quotient:**

So far we have discussed non-orientable symbols. The quotient rules can be defined for orientable symbols. The orientability of a symbol should not be confused with the usual meaning of an orietable surface. It is defined in the following definition.

**Definition 2.3:** *A symbol is said to be orientable if all of its faces are observable. A symbol is said to be non-orientable otherwise.*

For example a dice has six faces. A dice is said to be orientable if we are allowed to see all of its faces by flipping it over. A coin has two faces. However if a coin is pasted on a surface then only one side of it is observable. In this case the coin is said to be non-orientable. If the two faces of an orientable coin have same colors, then the faces are only distinguishable by their orientation.

**Definition 2.4:** *The faces of a symbol are said to be indistinguishable if they are only distinguishable by their orientations. Otherwise the faces are distinguishable.*

The faces of distinguishable symbols have different colors. We shall discuss **1**-axis case here. Some operations of orientable dots will be discussed in detail. A more general case will be considered elsewhere.

Define an operation $\oslash$ such that

$$\left(\mathbf{1}^\uparrow \oslash \mathbf{1}^\downarrow\right)_l = \left(\mathbf{1}^\uparrow \otimes \mathbf{1}_\downarrow\right)_l = \left(\overline{\mathbf{1}^\uparrow_\downarrow} \cdot {}^\uparrow_\downarrow \overline{\mathbf{1}^\uparrow_\downarrow}\right)_l$$

The strings of $\mathbf{1}^\uparrow_\downarrow$ to the left and right of $\bullet^\uparrow_\downarrow$ represent the left and right orientable arms respectively and $\bullet^\uparrow_\downarrow$ is the orientable dot that represents orientable shaking hands. The flipping of a face is explained below.

Two non-orientable dots can be pasted in many different ways. Let us first assign names to various non-orientable and distinguishable dots.

$\bullet_\downarrow$ represents a face − down dot oriented downward

$\bullet_\uparrow$ represents a face − up dot oriented downward

$\bullet^\uparrow$ represents a face − up dot oriented upward

$\bullet^\downarrow$ represents a face − down dot oriented upward

In the above nomenclatures, the color indices are omitted for brevity. For a coin, let face-up is the head. If the head is oriented upward then it is represented by $\bullet^\uparrow$ etc.

One can flip a face by finding its inverse

$$\left(\bullet^\uparrow\right)^{-1} = \bullet_\uparrow$$

$$\left(\bullet^\downarrow\right)^{-1} = \bullet_\downarrow$$

$$\left(\bullet_\downarrow\right)^{-1} = \bullet^\downarrow$$

$$\left(\bullet_\uparrow\right)^{-1} = \bullet^\uparrow$$

Let us establish various pasting schemes of the dots.

i) Faceless dots

$$\bullet^\downarrow_\uparrow = \bullet^\downarrow \otimes \bullet_\uparrow$$

$$\bullet^\uparrow_\uparrow = \bullet_\uparrow \otimes \bullet^\downarrow$$

Here the two dots are pasted in such a way that their faces are facing each other. The resultant dot appears faceless.

ii) Double face dots

$$\bullet^\uparrow \otimes \bullet_\downarrow = \bullet^\uparrow_\downarrow$$

The dots are pasted back to back. The resultant dot appears with two faces. For illustration purpose, we have considered only one double face dot.

iii)     Mixing

$$\bullet^\uparrow \otimes \bullet^\downarrow = \bullet^{\uparrow\downarrow}$$

The two dots are mixed up so that both are oriented upward. The resultant dot is also oriented upward. Similarly the other mixing can be obtained.

## 2.2 Extended Symbols and Substructure of a Dot

A dot is a symbol by definition-2.1 and secondly every symbol is connected by the same definition. One can assume any symbol as a dot at a different scale and so the symbols for that scale can be constructed. Consider **1 • 1** as shown in figure (2.5-a). The symbol in figure (2.5-a) is duplicated and pasted to the left and right of the original symbol. The extended symbol is as shown in figure (2.5-b). It should also be noted that colors of the duplicated symbols are reversed because the central box and the concatenated boxes should have opposite colors. The original symbol (as in figure (2.5-a) is considered to be a dot in the extended symbol (figure (2.5-b). The subscript numbers represented the scaling indices for the extended symbol. Similarly the new symbol can be considered as a dot at a larger. Continuing in this way, one gets a number which represents a fractal.

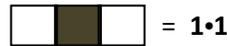 = **1•1**

Figure (2.5-a)

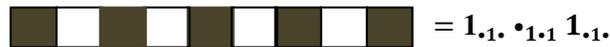 = $1_{\cdot 1\cdot} \bullet_{1\cdot 1} 1_{\cdot 1\cdot}$

Figure (2.5-b)

## 2.3. Broken Symbols

We have discussed symbols which obey definition-2.1. Broken symbols can be obtained by moving to definition-2.2. For illustration purpose, the following broken symbols are constructed.

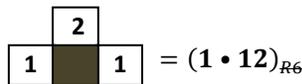 = $(1 \bullet 12)_{R6}$

Figure (2.6-a)

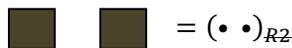 = $(\bullet\ \bullet)_{R2}$

Figure (2.6-b)

In figure (2.6-a), the symbol is broken along the **2**-axis. It violates rule R6 of definition-2.1. The subscript ~~R6~~ means that R6 is violated. If **1**-axis is extended to infinity on both sides of the dot and **2**-axis is only extended to infinity vertically above the dot then it is represented by $(\overline{1} \bullet \overline{12})_{R6}$. Similarly the symbol in figure (2.6-b) violates R2.

Let $m_{R1\,R2}$ be a broken number which violates R1 and R2, then

$$m_{R1\,R2} = m_{R2\,R1}$$

It is true because we do not care which rule is violated first.

## 3. Results and Discussion

In this section, the results are discussed qualitatively. For illustration purpose we only discuss applications of the symbols to dark matter.

Consider a two-dimensional symbol, $(1 \otimes 2)_2$, as shown in figure (3.1-a). Translate each axis by one unit. Definition-2.1 requires that every operation should be performed symmetrically otherwise we do not get a symbol. Symmetrical operation can obtained by duplicating each axis and separate them by same number of units both horizontally and vertically as shown in figure (3.1-b). The resultant symbol satisfies all properties of definition-2.1, therefore it is a symbol.

The interesting region is the central island **D**, which neither belongs to **1**-axes nor **2**-axes. By careful inspection, one can observe that the boundaries of **D** are closed, that is **D** is connected collection of straight lines. Therefore **D** is a dot. Hence **D** is a part of the expended symbol.

Let the symbol in figure (3.1-a) represents a two-dimensional universe and the dark dot represent a particle. When the symbol (the universe) expends, it creates four particles and a central island **D**. We interpret **D** to be the dark matter counterpart of the space. If the symbol is further allowed to expend, then no more particles are created but the central region **D** grows up. Said differently the dark matter increases as the universe expends.

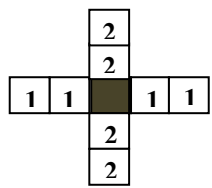 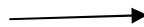 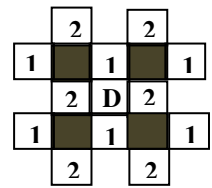

Figure (3.1-a)                                    Figure (3.1-b)

## 4. Conclusion

In this work, the symbols for flat space have been constructed. The symbols are then coded into '*many dots numbers* (MDN)'. Two kinds of symbols have been defined. One of them has been named '*symbols*' and the other as '*broken symbols*'. The symbols and the broken symbols may correspond to symmetries and broken symmetries respectively. The symbol algebra developed above is true for all scales because the size of a dot is arbitrary. As a consequence, one can assume an atom as a dot and at much larger scale the entire universe can be a dot. The colors and faces of the dots are very crucial. If the dots are interpreted as particles, then the colors may represent mass and the faces may correspond to the spin of the particles. Moreover MDN are abstract numbers, they may have application in any physical problems. More interesting results are expected for the symbols of other manifolds such as sphere, cylinder etc. These possibilities will be discussed in the future work.